\documentclass[conference]{IEEEtran}
\IEEEoverridecommandlockouts

\usepackage{cite}
\usepackage{amsmath,amssymb,amsfonts}
\usepackage{algorithmic}
\usepackage[acronym]{glossaries}
\usepackage{graphicx}
\usepackage{hyperref}
\usepackage{textcomp}
\usepackage{authblk}
\usepackage{xcolor}
\def\BibTeX{{\rm B\kern-.05em{\sc i\kern-.025em b}\kern-.08em
    T\kern-.1667em\lower.7ex\hbox{E}\kern-.125emX}}

\newacronym{rss}{RSS}{received signal strength}
\newacronym{ple}{PLE}{path loss exponent}
\newacronym{awgn}{AWGN}{additive white Gaussian noise}
\newacronym{aoa}{AoA}{angle of arrival}
\newacronym{ml}{ML}{maximum likelihood}
\newacronym{wls}{WLS}{weighted least squares}
\newacronym{pdf}{PDF}{probability distribution function}
\newacronym{mlp}{MLP}{multilayer perceptron}
\newacronym{mse}{MSE}{mean squared error}
\newacronym{ls}{LS}{least square}
\newacronym{5g}{5G}{fifth-generation wireless system}
\newacronym{6g}{6G}{sixth-generation wireless system}
\newacronym{embb}{eMBB}{enhanced mobile broadband}
\newacronym{urllc}{URLLC}{ultra-reliable low-latency communications}
\newacronym{mmtc}{mMTC}{massive machine-type communications}
\newacronym{iot}{IoT}{Internet of Things}
\newacronym{v2x}{V2X}{vehicle-to-everything communication}
\newacronym{uav}{UAV}{unmanned aerial vehicle}
\newacronym{mimo}{MIMO}{multiple-input multiple-output}
\newacronym{dl}{DL}{deep learning}
\newacronym{ai}{AI}{aritifical intelegence}
\newacronym{rmse}{RMSE}{root mean square error}
\begin{document}

\title{Impact of Preprocessing on Neural Network–Based RSS/AoA Positioning\\
{}
\thanks{This work was conducted within the MiFuture project, which has received funding from the European Union’s Horizon Europe (HE) Marie
	
	Skłodowska-Curie Actions MiFuture HORIZON-MSCA2022-DN-01, under Grant Agreement number 101119643 and YAHYA/6G HORIZON-MSCA-2022-PF-01, under Grant Agreement number 101109435. It was partially supported by FCT/MECI through national funds and when applicable co-funded EU funds under UID/50008: Instituto de Telecomunicações.}
}

\author[1,2]{Omid Abbassi Aghda}
\author[3,4]{Slavisa Tomic}
\author[1,5]{Oussama Ben Haj Belkacem} 
\author[1,2]{João Guerreiro} 
\author[1,6]{Nuno Souto} 
\author[2,7]{\\Michal Szczachor} 
\author[1,2]{Rui Dinis} \affil[1]{Instituto de Telecomunicações, Lisboa, Portugal} \affil[2]{Universidade Nova de Lisboa, Monte da Caparica, 2829-516 Caparica, Portugal} 
\affil[3]{UNINOVA-CTS – Center of Technology and Systems, NOVA School of Science and Technology, 2829-516 Caparica, Portugal}
\affil[4]{COPELABS, ECATI, Lusófona University, 1749-024 Lisbon, Portugal}
\affil[5]{Innov’Com Laboratory, Sup’Com, University of Carthage, Tunis 1054, Tunisia} 
\affil[6]{ ISCTE-Instituto Universitário de Lisboa,
	1649-026 Lisbon, Portugal}
\affil[7]{Nokia, Wroclaw, Poland}

\maketitle

\begin{abstract}
Hybrid received signal strength (RSS)/angle of arrival (AoA)-based positioning offers low-cost distance estimation and high-resolution angular measurements. Still, it comes at a cost of  inherent nonlinearities, geometry-dependent noise, and suboptimal weighting in conventional linear estimators that might limit accuracy. In this paper, we propose a neural network–based approach using a multilayer perceptron (MLP) to directly map RSS/AoA measurements to 3D positions, capturing nonlinear relationships that are difficult to model with traditional methods. We evaluate the impact of input representation by comparing networks trained on raw measurements versus preprocessed features derived from a linearization method. Simulation results show that the learning-based approach consistently outperforms existing linear methods under RSS noise across all noise levels, and matches or surpasses state-of-the-art performance under increasing AoA noise. Furthermore, preprocessing measurements using the linearization method provides a clear advantage over raw data, demonstrating the benefit of geometry-aware feature extraction. 
\end{abstract}

\begin{IEEEkeywords}
Deep learning, 3D localization, received signal strength (RSS), angle of arrival (AoA), weighted least squares (WLS)
\end{IEEEkeywords}

\section{Introduction}
\Gls{5g} wireless systems are designed to support a diverse set of use cases, including \gls{embb}, \gls{urllc}, and \gls{mmtc}. Beyond data transmission, these use cases increasingly rely on accurate situational awareness to enable higher-layer applications such as \gls{iot} services, \gls{v2x} communication, \gls{uav}s, and autonomous vehicular systems. In such applications, awareness of the position of communicating devices or targets, such as vehicles or drones, is essential for safe and efficient operation. Consequently, wireless positioning has emerged as a key enabling technology that must be addressed alongside communication performance. Looking ahead, it is widely anticipated that in \gls{6g} systems, positioning and communication will be even more tightly integrated, with location information actively supporting communication, sensing, and control functions \cite{Italiano10644093,Yang10592075}.

One common approach for positioning is to use \gls{rss} and \gls{aoa} measurements collected from sensors. \Gls{rss} information is attractive due to its low cost, while angular measurements can be obtained efficiently in multi-antenna \gls{mimo} systems. A well-studied research direction is to transform the inherently nonlinear positioning problem based on \gls{rss} and \gls{aoa} into a linear form to enable closed-form solutions. For example, in \cite{Tomic2016Closed}, the authors approximate the nonlinear model using a Taylor series expansion and a Cartesian-to-spherical transformation, resulting in a linearized system that can be solved via \gls{wls}. Similarly, the approach in \cite{Kegen4350292} applies an alternative linearization method and uses a standard \gls{ls} estimator to solve the resulting system.
	Subsequent work in \cite{TomicSlavisa8607068} extends this approach by adding constraints to the \gls{wls} cost function, reformulating it as a generalized trust region subproblem (GTRS), and solving it efficiently using a bisection procedure. Despite their computational efficiency, these methods remain limited in practice. \Gls{rss}- and \gls{aoa}-based positioning suffers from model nonlinearities, multipath propagation, and noise interference, which reduce accuracy in real-world deployments. Moreover, linearization based on Taylor approximations introduces errors, and the resulting \gls{wls} solution is generally suboptimal under realistic noise conditions \cite{yang2025bluetooth,kang2020hybrid}. In addition, the weighting matrices used in \gls{wls} are typically heuristic or approximate, and non-optimal weighting can further degrade performance, especially when measurement noise is heterogeneous across anchors \cite{kang2020hybrid}.

The use of \gls{ai} techniques is widespread in positioning problems. In particular, \gls{dl}-based models are well suited for capturing nonlinear relationships and can effectively model the nonlinearities inherent in hybrid \gls{rss}/\gls{aoa} positioning \cite{UniversalApproximation2025}. Moreover, the positioning problem is inherently geometry-dependent, and preprocessing has a significant impact on the performance of \gls{dl}-based systems \cite{burghal2020comprehensive}. Therefore, the system structure and input representation should be explicitly considered rather than relying solely on raw measurements. Consequently, analyzing whether to use preprocessed geometric features or raw measurements is essential to determine which approach enables more effective learning.

In this paper, we propose a neural network-based approach to address the nonlinearities inherent in hybrid \gls{rss}/\gls{aoa} positioning, as studied in \cite{Tomic2016Closed,TomicSlavisa8607068}. Specifically, we design and train a \gls{mlp} network to learn the mapping from measurements to 3D positions, capturing nonlinear relationships that are difficult to model with conventional linear estimators. In addition, we evaluate the impact of input representation by comparing networks trained on raw measurements versus preprocessed features derived from the linearization method in \cite{Tomic2016Closed}, allowing us to quantify how geometry-aware preprocessing can enhance learning performance.

Although \gls{wls} approach provides a computationally efficient estimator, its performance is limited in practice due to linearization errors, noise propagation, and potentially suboptimal weighting. By leveraging \gls{mlp}, our approach can implicitly compensate for these limitations, learning a mapping that mitigates nonlinearities, weighting biases, and geometry-dependent errors, resulting in more robust positioning. To validate our method, we plot the \gls{rmse} versus noise variance for both \gls{rss} and \gls{aoa} measurements. The results show that the \gls{mlp}-based approach consistently outperforms the methods in \cite{Tomic2016Closed,Kegen4350292,TomicSlavisa8607068} in the presence of \gls{rss} noise across all noise levels. For increasing \gls{aoa} noise, the proposed method yields lower RMSE than \cite{Kegen4350292,TomicSlavisa8607068}, while achieving RMSE performance comparable to \cite{Tomic2016Closed}. Furthermore, using the same network architecture, preprocessing the measurements via the linearization method provides a clear advantage over training with raw data, demonstrating the benefit of geometry-aware feature extraction.

\section{Problem Formulation}

\begin{figure}[t]
	\centering
	\includegraphics[width=0.8\linewidth]{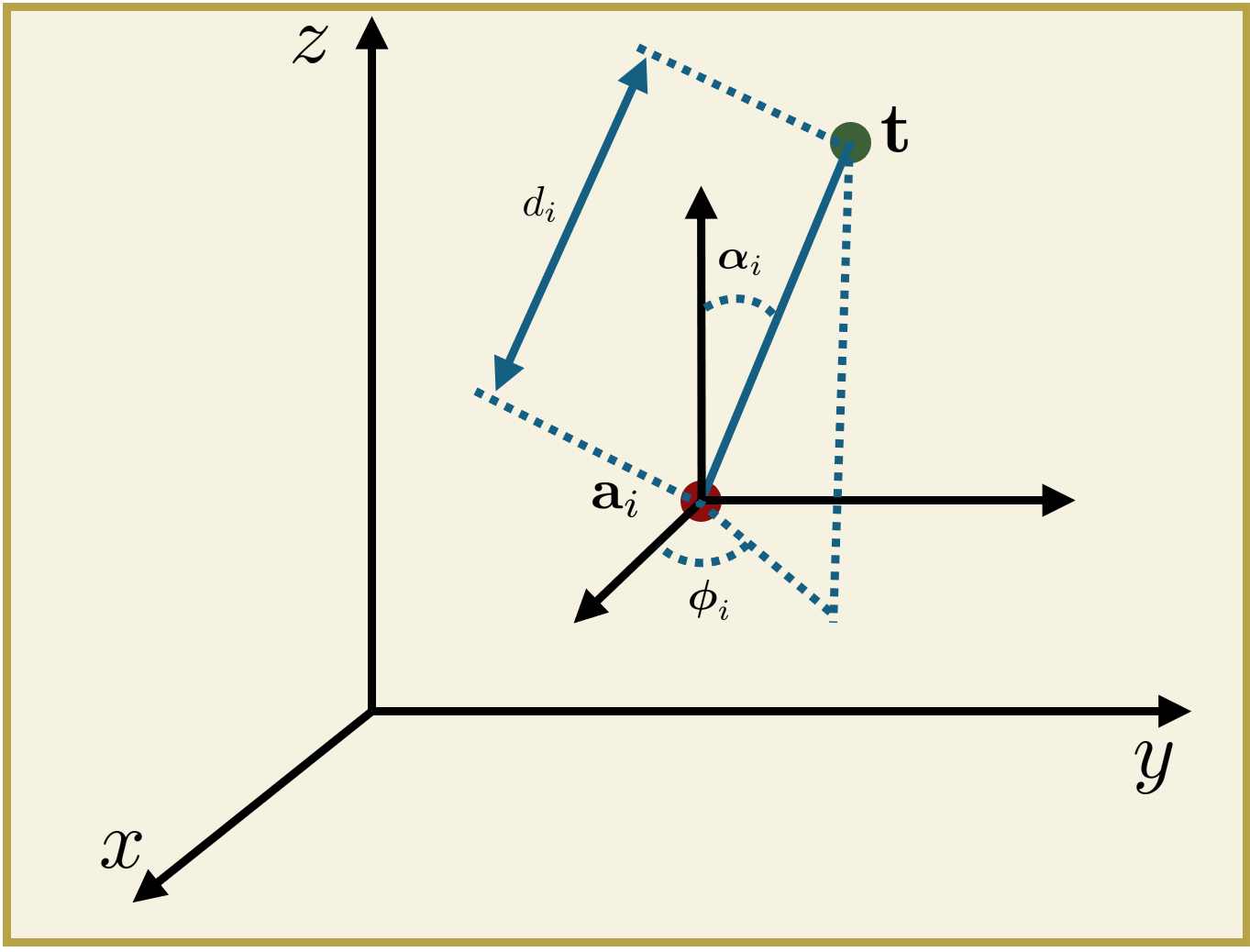}
	\caption{3d system model}
	\label{fig:system model}
\end{figure}

We consider the problem of estimating the unknown three-dimensional location of a target, denoted by $\mathbf{t} \in \mathbb{R}^3$, which is assumed to lie within a bounded region of interest modeled as a box of size $B$. The positions of $N$ anchors are known and represented by $\mathbf{a}_i \in \mathbb{R}^3$, $i = 1,\dots,N$. Specifically, the target and anchor coordinates are given by $\mathbf{t} = [t_x, t_y, t_z]^T$ and $\mathbf{a}_i = [a_{ix}, a_{iy}, a_{iz}]^T$, and the system geometry is illustrated in Fig.~\ref{fig:system model}. For each anchor–target pair, $d_i$, $\phi_i$, and $\alpha_i$ denote the true distance, azimuth angle, and elevation angle, respectively.

The distance $d_i$, which can be estimated from the \gls{rss} at the $i$-th anchor is modeled as
\begin{align}
	P_i = P_0 - 10 \, \gamma \, \log_{10} \frac{d_i}{d_0} + n_i,
	\label{eq:rss}
\end{align}
where $P_0$ is the reference power at distance $d_0$  (usually $d_0 = 1$), $\gamma$ is the path-loss exponent, and $n_i \sim \mathcal{N}(0,\sigma_{n_i}^2)$ represents the log-normal shadowing noise in the logarithmic domain.

The \gls{aoa} measurements are obtained using a directional antenna array and are modeled as
\begin{align}
	\phi_i &= \arctan \left( \frac{t_y - a_{iy}}{t_x - a_{ix}} \right) + m_i,
	\label{eq:azi}
\end{align}
\begin{align}
	\alpha_i &= \arccos \left( \frac{t_z - a_{iz}}{\lVert \mathbf{t} - \mathbf{a}_i \rVert} \right) + \nu_i,
	\label{eq:ele}
\end{align}
where $m_i \sim \mathcal{N}(0, \sigma_{m_i}^2)$ and $\nu_i \sim \mathcal{N}(0, \sigma_{\nu_i}^2)$ denote azimuth and elevation measurement errors, respectively.

To estimate the target position, one can form the conditional probability density function (PDF) of the measurements given $\mathbf{t}$. Let the observation vector be $\boldsymbol{\theta} = [\mathbf{p}^T, \boldsymbol{\phi}^T, \boldsymbol{\alpha}^T]^T \in \mathbb{R}^{3N}$, where $\mathbf{p} = [P_1, \dots, P_N]^T$, $\boldsymbol{\phi} = [\phi_1, \dots, \phi_N]^T$, and $\boldsymbol{\alpha} = [\alpha_1, \dots, \alpha_N]^T$. Assuming independent Gaussian measurement noise, the likelihood function can be written as
\begin{align}
	P(\boldsymbol{\theta}\mid\mathbf{t})
	&= \prod_{j=1}^{3}\prod_{i=1}^{N}
	\frac{1}{\sqrt{2\pi\sigma_{i+(j-1)N}^2}} \nonumber\\
	&\quad \times
	\exp\!\left\{
	- \frac{\big(\theta_{i+(j-1)N}
		- {f}_{i+(j-1)N}(\mathbf{t})\big)^2}
	{2\sigma_{i+(j-1)N}^2}
	\right\},
	\label{eq:pdf_of_estimation}
\end{align}

where
\begin{align}
	\mathbf{f}(\mathbf{t})
	&= \big[ {f}_1(\mathbf{t}), \ldots, {f}_{3N}(\mathbf{t}) \big] \\
	&=
	\begin{aligned}[t]
		\big[
		& P_0 - 10\gamma \log_{10}\!\left(\dfrac{d_1}{d_0}\right), \cdots ,
		P_0 - 10\gamma \log_{10}\!\left(\dfrac{d_N}{d_0}\right), \\
& \arctan\!\left(\dfrac{t_y - a_{1y}}{t_x - a_{1x}}\right), \cdots,
		\arctan\!\left(\dfrac{t_y - a_{Ny}}{t_x - a_{Nx}}\right), \\
		& \arccos\!\left(\dfrac{t_z - a_{1z}}{\lVert \mathbf{t} - \mathbf{a}_1 \rVert}\right), \cdots,
		\arccos\!\left(\dfrac{t_z - a_{Nz}}{\lVert \mathbf{t} - \mathbf{a}_N \rVert}\right)
		\big]
	\end{aligned}.
\end{align}

Directly maximizing this likelihood is challenging due to the nonconvexity of the resulting optimization problem.

An alternative, suboptimal approach is the \gls{wls} solution proposed in \cite{Tomic2016Closed}, which approximates the nonlinear model as a linear system:
\begin{align}
	\min_{\mathbf{t}} \quad \lVert \mathbf{W} (\mathbf{A} \mathbf{t} - \mathbf{b}) \rVert,
	\label{eq:wls_problem}
\end{align}
with closed-form solution
\begin{align}
	\hat{\mathbf{t}} = (\mathbf{A}^T \mathbf{W}^T \mathbf{W} \mathbf{A})^{-1} (\mathbf{A}^T \mathbf{W}^T \mathbf{W} \mathbf{b}),
\end{align}
where the matrices have dimensions $\mathbf{A} \in \mathbb{R}^{3N \times 3}$, $\mathbf{b} \in \mathbb{R}^{3N \times 1}$, and $\mathbf{W} \in \mathbb{R}^{3N \times 3N}$, with detailed definitions provided in the Appendix.

\begin{figure}[t]
	\centering
	\includegraphics[width=1\linewidth]{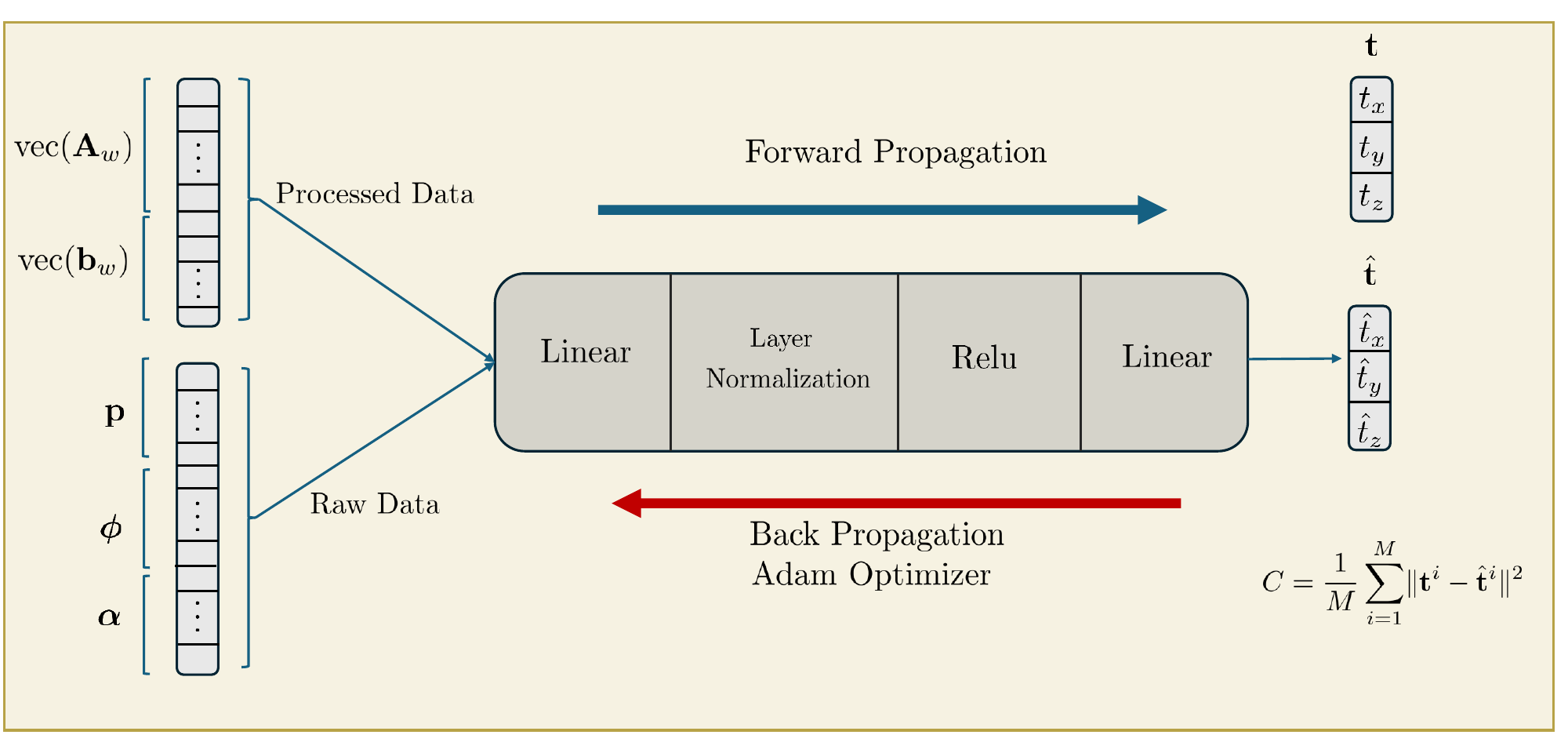}
	\caption{Illustration of the MLP-based positioning network. The model can operate on two types of inputs: preprocessed features derived from the WLS formulation ($\text{vec}(\mathbf{A}_w), \text{vec}(\mathbf{b}_w)$) or raw measurements ($\mathbf{p}, \boldsymbol{\phi}, \boldsymbol{\alpha}$). The network consists of a linear layer, layer normalization, ReLU activation, and a final linear layer to predict the 3D target position $\hat{\mathbf{t}} = [\hat{t}_x, \hat{t}_y, \hat{t}_z]^T$. Training is performed using the MSE loss with the Adam optimizer.}
	\label{fig:system setup}
\end{figure}

Although \gls{wls} provides a computationally efficient estimator, it suffers from several limitations in practical positioning systems. Its performance relies on assumptions such as accurate linearization, Gaussian and independent noise, and correctly specified weighting, which are only approximately satisfied in practice. In particular, linearization assumes small measurement noise for all observations, but performance is much more sensitive to \gls{aoa} noise than \gls{rss} noise. Small angular errors in \gls{aoa} measurements can lead to large position errors due to the geometric relationship between angles and location \cite{alsharif2023pipe}. Moreover, linearized formulations of hybrid \gls{rss}/\gls{aoa} models do not produce strictly independent Gaussian errors, since the linearization propagates measurement noise nonlinearly \cite{kang2020hybrid}.

The measurement noise is also heterogeneous across anchors, meaning that different anchors may have different variances \cite{kang2020hybrid}. The weighting matrix proposed in \cite{Tomic2016Closed}, while reasonable, is not guaranteed to be optimal; if it is uniform or estimated incorrectly, the \gls{wls} solution can be biased. These factors collectively limit the accuracy of \gls{wls}, particularly in scenarios with high \gls{aoa} noise.

To address these limitations, we propose using \gls{mlp} to learn a robust mapping from the \gls{wls} input parameters to the target position. Our approach can implicitly account for nonlinearities, heterogeneous noise, and suboptimal weighting, leading to nearly constant performance across different SNR levels, unlike conventional \gls{wls}.

\section{Deep learning based positioning}

To address the limitations of \gls{wls}, including nonlinear noise propagation, heterogeneous measurement variances, and sensitivity to weighting accuracy, we consider two \gls{mlp}-based positioning approaches within the same \gls{mlp} architecture. The first approach leverages the model-based preprocessing proposed in \cite{Tomic2016Closed}, while the second directly operates on raw measurements. This allows us to isolate and evaluate the impact of preprocessing on learning performance, while maintaining an identical network structure and training procedure.
Figure~\ref{fig:system setup} shows the \gls{mlp} network architecture for the cases of raw and preprocessed inputs.

\subsection{Learning with preprocessed measurements}

In the first approach, the \gls{mlp} is trained using the linearized system parameters derived from the \gls{wls} formulation. Specifically, the matrices $\mathbf{A}$, $\mathbf{b}$, and the weighting matrix $\mathbf{W}$ are used as input features, while the true target position $\mathbf{t}$ serves as the label.

Each training sample is formed by preprocessing the inputs into a single feature vector
\begin{align}
	\mathbf{x}^i = [\text{vec}(\mathbf{A}_w)^T, \text{vec}(\mathbf{b}_w)^T]^T \in \mathbb{R}^{12N \times 1},
\end{align}
where $\mathbf{A}_w = \mathbf{W}\mathbf{A} \in \mathbb{R}^{3N \times 3}$ and $\mathbf{b}_w = \mathbf{W}\mathbf{b} \in \mathbb{R}^{3N \times 1}$.

The training dataset is constructed as
\[
\mathbf{X} = [\mathbf{x}^1, \ldots, \mathbf{x}^M] \in \mathbb{R}^{12N \times M}, \quad
\mathbf{Y} = [\mathbf{t}^1, \ldots, \mathbf{t}^M] \in \mathbb{R}^{3 \times M},
\]
where $M$ denotes the number of training samples and $\mathbf{t}^i$ is the target position corresponding to $\mathbf{x}^i$.

The \gls{mlp} is trained to minimize the \gls{mse}
\begin{align}
	C = \frac{1}{M} \sum_{i=1}^{M} \lVert \mathbf{t}^i - \hat{\mathbf{t}}^i \rVert^2,
\end{align}
using  Adam optimizer.

\subsection{Learning with raw measurements}

To evaluate the role of model-based preprocessing, we also train an \gls{mlp} using the raw measurement vector as input. In this case, the input matrix is defined as
\[
\mathbf{X}_{\text{raw}} = [\boldsymbol{\theta}^1, \ldots, \boldsymbol{\theta}^M] \in \mathbb{R}^{3N \times M},
\]
where $\boldsymbol{\theta}^i = [\mathbf{p}^{iT}, \boldsymbol{\phi}^{iT}, \boldsymbol{\alpha}^{iT}]^T$ contains the raw \gls{rss} and \gls{aoa} measurements for the $i$-th sample. The output labels are identical to those used in the preprocessed case, i.e.,
$
\mathbf{Y} = [\mathbf{t}^1, \ldots, \mathbf{t}^M].
$

Apart from the input representation, only the input dimensionality differs between the two approaches, while the network architecture, training procedure, and loss function remain identical.
 This setup enables a fair comparison between learning from raw measurements and learning from model-informed features derived from the \gls{wls} formulation.

\section{Simulation Results}

Simulations are performed for a target constrained within a three-dimensional box of size $B = 15$, with $N = 4$ anchors. The path-loss exponent $\gamma$ is assumed to vary between $2.2$ and $2.8$, while the receiver uses a fixed value of $\gamma = 2.5$. Other system parameters are $P_0 = -10$~dBm and $d_0 = 1$~m. 
The \gls{wls} and \gls{ls} solutions are evaluated over $10{,}000$ Monte Carlo iterations for each noise scenario. For the \gls{mlp}-based estimators, the dataset contains $100{,}000$ samples, of which $75\%$ are used for training, $15\%$ for validation, and $10\%$ for testing. The network is trained for 300 epochs using a learning rate of 0.01. The network is trained on all SNR levels and scenarios, while testing is performed on specific noise levels to generate the performance curves. The network architecture and training setup are summarized\footnote{The dataset and trained networks are available at \url{https://your-link-here}.}
 in Table~\ref{tab}. The \gls{mlp} architecture was selected through an empirical design process. An initially overparameterized network was considered to ensure sufficient modeling capacity, after which the number of layers and neurons was progressively reduced to improve generalization. Batch normalization was found ineffective in this setting and was therefore not adopted, while layer normalization was employed to stabilize training under varying noise conditions. The final architecture provides a favorable trade-off between estimation accuracy and model complexity.

To evaluate performance, the \gls{rmse} of the estimated target positions is computed as
\begin{align}
	\text{\gls{rmse}} = \sqrt{\frac{1}{M_c} \sum_{i=1}^{M_c} \lVert \mathbf{t}_i - \hat{\mathbf{t}}_i \rVert^2},
\end{align}
where $M_c$ denotes the number of Monte Carlo iterations, $\mathbf{t}_i$ is the true target position, and $\hat{\mathbf{t}}_i$ is the corresponding estimate. 
For the \gls{mlp}-based estimators, the input features are normalized to have zero mean and unit variance, ensuring that all features have comparable statistical scales. This preprocessing step facilitates stable and efficient network training across different types of measurements.  
 The methods considered as the benchmark in this study are named as follows: the baseline methods from \cite{Tomic2016Closed}, \cite{Kegen4350292}, and \cite{TomicSlavisa8607068} are referred to as WLS, LS, and SR-WLS, respectively, while the proposed deep learning-based approach is referred to as "MLP-Processed Data" and "MLP-Raw Data".
 
\begin{figure}[t]
	\centering
	\includegraphics[width=0.9\linewidth]{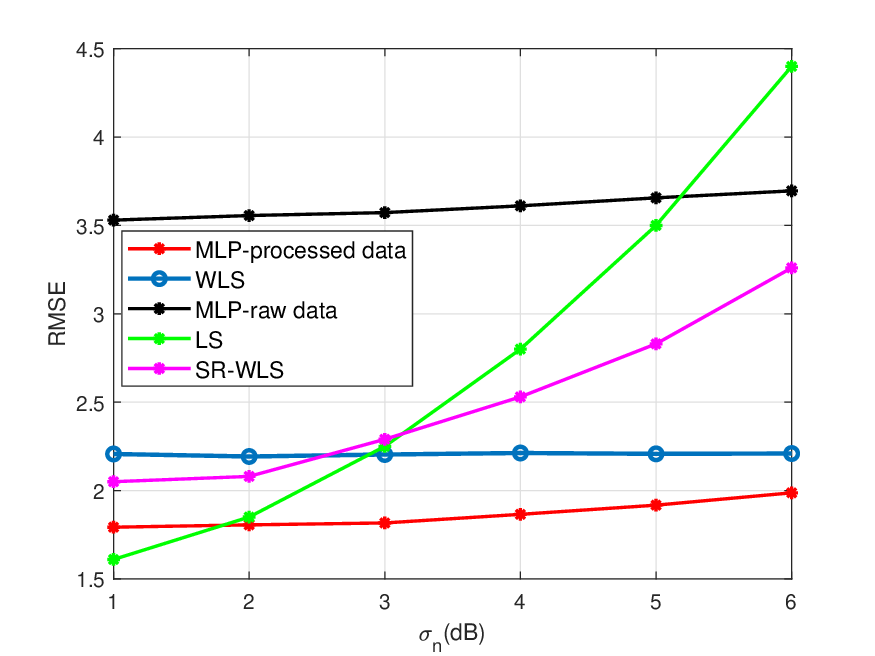}
	\caption{RMSE of the target position versus \gls{rss} noise standard deviation $\sigma_{n_i}$.}
	\label{fig:rmse_rss}
\end{figure}

 Figure~\ref{fig:rmse_rss} shows the \gls{rmse} as a function of the \gls{rss} noise standard deviation $\sigma_n$. It can be observed that the \gls{mlp}-based estimator trained with preprocessed measurements using the linearization method in \cite{Tomic2016Closed}  consistently outperforms all other methods, including WLS, SR-WLS, and LS. In contrast, training the same network on raw measurements does not provide any significant improvement over the conventional estimators. 
 
\begin{figure}[t]
	\centering
	\includegraphics[width=0.9\linewidth]{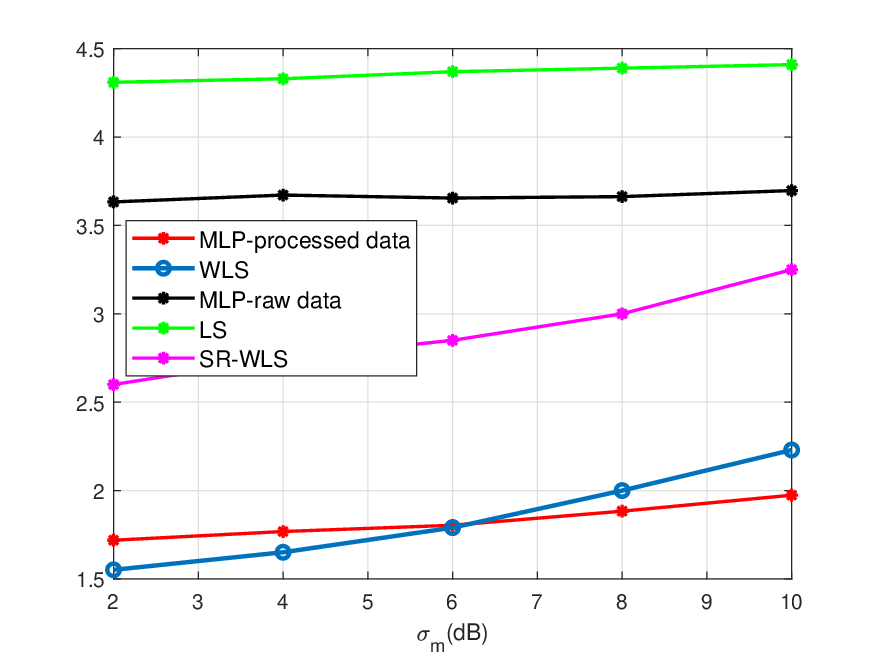}
	\caption{RMSE of the target position versus azimuth noise standard deviation $\sigma_{m_i}$.}
	\label{fig:rmse_azimuth}
\end{figure}

\begin{figure}[t]
	\centering
	\includegraphics[width=0.9\linewidth]{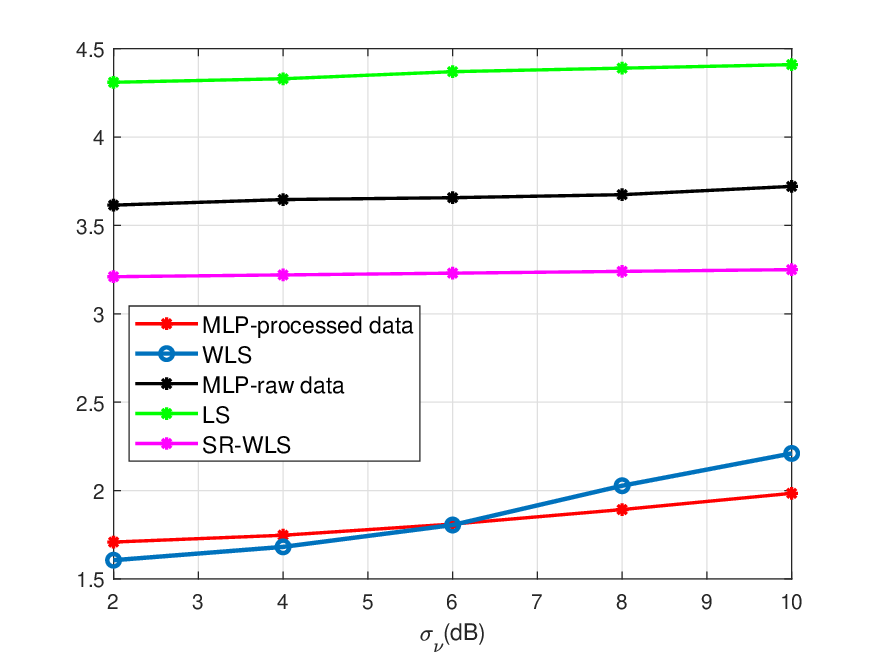}
	\caption{RMSE of the target position versus elevation noise standard deviation $\sigma_{\nu_i}$.}
	\label{fig:rmse_elevation}
\end{figure}

 Figures~\ref{fig:rmse_azimuth} and \ref{fig:rmse_elevation} depict the \gls{rmse} versus azimuth and elevation noise, $\sigma_m$ and $\sigma_\nu$, respectively. The results for both angular measurements are similar: \gls{mlp}-preprocessed data  achieves better performance than SR-WLS and LS, while its performance is comparable to WLS. Furthermore, the advantage of using preprocessed measurements persists across all noise levels, demonstrating that the proposed approach provides robust performance under a wide range of noise conditions.

\begin{table}[t]
	\centering
	\caption{DNN Architecture and Parameters}
	\begin{tabular}{|c|c|c|c|}
		\hline
		\textbf{Layer} & \textbf{Type} & \textbf{Output Size} & \textbf{Notes / Parameters} \\
		\hline
		1 & Linear & 128 & Input: $\mathbf{x}^i$ / $\boldsymbol{\theta}^i$  \\
		2 & LayerNorm & 128 & Normalizes features across layer \\
		3 & ReLU & 128 & Activation function \\
		4 & Linear & 3 & Output: $\hat{\mathbf{t}}_i$ \\
		\hline
	\end{tabular}
	\label{tab}
\end{table}

\section{conclusion}
In this paper, we proposed a \gls{mlp}-based approach for hybrid \gls{rss}/\gls{aoa} positioning, capturing nonlinearities that limit conventional linear estimators. We showed that preprocessing measurements using a linearization method significantly improves the performance of the \gls{mlp} network compared to raw data. Simulation results demonstrated that the proposed method consistently outperforms existing WLS, SR-WLS, and LS approaches under RSS and angular noise. These findings highlight that deep learning with geometry-aware preprocessing enables robust and accurate positioning across a wide range of noise conditions.

\bibliography{vtc2026references}
\bibliographystyle{IEEEtran}

\appendix
\section{Linearizing the Positioning Problem}
Assuming the noise is negligible, \eqref{eq:rss}, \eqref{eq:azi}, and \eqref{eq:ele} can be approximated as

\begin{equation}
	\lambda_i \|\mathbf{t} - \mathbf{a}_i\| \approx \eta d_0, \quad i = 1, \dots, N
	\label{eq:rss no noise}
\end{equation}
\vspace{-5mm}
\begin{equation}
	\mathbf{c}_i^T (\mathbf{t} - \mathbf{a}_i) \approx 0, \quad i = 1, \dots, N
	\label{eq:azi no noise}
\end{equation}
\vspace{-5mm}
\begin{equation}
	\mathbf{k}^T (\mathbf{t} - \mathbf{a}_i) \approx \|\mathbf{x} - \mathbf{a}\| \cos(\alpha_i), \quad i = 1, \dots, N
	\label{eq:ele no noise}
\end{equation}
where
	$\lambda_i = 10^{P_i / 10\gamma}$, $
	\eta = 10^{P_0 / 10\gamma}$, 
	$\mathbf{c}_i = [-\sin(\phi_i), \cos(\phi_i), 0]^T$, 
	$\mathbf{k} = [0,0,1]^T$.

If each  $\mathbf{t}- \mathbf{a}_i$ in the spherical domain is represented as 
\begin{align}
	\mathbf{t}-\mathbf{a}_i = r_i \mathbf{u}_i,
\end{align}
where $r_i > 0$ is the distance from the origin and $\lVert \mathbf{u}_i \rVert = 1$ is the directional unit vector obtained from the \gls{aoa} as
\begin{align}
	\mathbf{u}_i = [\cos(\phi_i)\sin(\alpha_i), \, \sin(\phi_i)\sin(\alpha_i), \, \cos(\alpha_i)]^T,
\end{align}
then, multiplying \eqref{eq:rss no noise}, \eqref{eq:azi no noise}, and \eqref{eq:ele no noise} by $\mathbf{u}_i^T \mathbf{u}_i$ gives

\begin{align}
	\lambda_i \mathbf{u}_i^T r_i \mathbf{u}_i 
	&\approx \eta d_0 
	\Leftrightarrow 
	\lambda_i \mathbf{u}_i^T (\mathbf{t} - \mathbf{a}_i) 
	\approx \eta d_0,
	\label{eq:lambda_r} \\
	\mathbf{k}^T r_i \mathbf{u}_i 
	&\approx \mathbf{u}_i^T r_i \mathbf{u}_i \cos(\alpha_i)
	\Leftrightarrow  
	(\cos(\alpha_i)\mathbf{u}_i - \mathbf{k})^T (\mathbf{t} - \mathbf{a}_i) 
	\approx 0,
	\label{eq:ele_sph}
\end{align}

From \eqref{eq:azi no noise}, \eqref{eq:lambda_r}, and \eqref{eq:ele_sph}, the matrices $\mathbf{A}$ and $\mathbf{b}$ can be directly constructed as
\begin{align}
	\mathbf{A} &=
	\begin{bmatrix}
		\vdots \\
		\lambda_i \mathbf{u}_i^{T} \\
		\vdots \\
		\mathbf{c}_i^{T} \\
		\vdots \\
		\left( \cos(\alpha_i)\mathbf{u}_i - \mathbf{k} \right)^{T} \\
		\vdots
	\end{bmatrix},
	\qquad
	\mathbf{b} =
	\begin{bmatrix}
		\vdots \\
		\lambda_i \mathbf{u}_i^{T} \mathbf{a}_i + \eta d_0 \\
		\vdots \\
		\mathbf{c}_i^{T} \mathbf{a}_i \\
		\vdots \\
		\left( \cos(\alpha_i)\mathbf{u}_i - \mathbf{k} \right)^{T} \mathbf{a}_i \\
		\vdots
	\end{bmatrix}.
\end{align}

The weighting matrix is constructed as $\mathbf{W} = \mathbf{I}_3 \otimes \mathbf{w}$, where $\mathbf{I}_3$ denotes the $3 \times 3$ identity matrix and $\mathbf{w} \in \mathbb{R}^{N \times 1}$ contains the weighting coefficients $w_i = 1 - \frac{\hat{d}_i}{\sum_{j=1}^{N} \hat{d}_j}$, with $\hat{d}_i = d_0 10^{\frac{P_0 - P_i}{10\gamma}}$. The symbol $\otimes$ denotes Kronecker  product.


\end{document}